\documentclass[12pt]{article}
\usepackage{longtable,supertabular}
\usepackage{amsmath}
\usepackage{amssymb}
\usepackage{latexsym}
\usepackage{cite,a4wide,color}
\usepackage{multirow}
\usepackage{ntheorem}
\usepackage[square,numbers]{natbib}

\theoremstyle{remark}

\title{\bf Cyclic and Negacyclic Codes with Optimal and Best Known Minimum Distances}
\author{Hao Chen and Yanan Wu\thanks{Hao Chen is with the College of Information Science and Technology/Cyber Security, Jinan University, Guangzhou, Guangdong Province, 510632, China, haochen@jnu.edu.cn. Yanan Wu is with Academy of Mathematics and Systems Science, Chinese Academy of Sciences, Beijing 100190, China, yanan.wu@aliyun.com. This research was supported by NSFC Grant 62032009.}}

\begin{document}
	
	\maketitle
	\begin{abstract}
In this paper, we construct infinitely many families of distance-optimal binary BCH codes with the minimum distance $6$ and a new infinite family of distance-optimal quaternary BCH codes with the minimum distance $4$.  We also construct several infinite families of cyclic and negacyclic BCH codes over ${\bf F}_2$, ${\bf F}_3$, ${\bf F}_4$, ${\bf F}_5$, ${\bf F}_7$ and ${\bf F}_9$ with good parameters $n,\,k,\,d$, such that the maximal possible minimum distance $d_{max}$ of a linear $[n, k]_q$ code is at most $d_{max} \leq d+8$.  Many codes in these families have optimal or best known minimum distances. $145$ optimal or best known codes are constructed as cyclic codes, negacyclic codes, their shortening codes and punctured codes. Several infinite families of rate $\frac{1}{2}$ negacyclic $[n, \frac{n+1}{2}, d]_q$ codes or $[n, \frac{n}{2}, d]_q$ codes, such that their minimum distances satisfy $d\geq \frac{cn}{\log_q n}$, where $c$ is a positive constant, are also constructed. These are first several families of such negacyclic codes.\\

\end{abstract}

\newpage
	
\section{Introduction}

\subsection{Preliminaries}

The Hamming weight $wt({\bf a})$ of a vector ${\bf a}=(a_0, \ldots, a_{n-1}) \in {\bf F}_q^n$ is the cardinality of its support, $$supp({\bf a})=\{i: a_i \neq 0\}.$$ The Hamming distance $d({\bf a}, {\bf b})$ between two vectors ${\bf a}$ and ${\bf b}$ is $d({\bf a}, {\bf b})=wt({\bf a}-{\bf b})$. Then ${\bf F}_q^n$ is a finite Hamming metric space. For a code ${\bf C} \subset {\bf F}_q^n$, its Hamming distance $$d=\min_{{\bf a} \neq {\bf b}} \{d({\bf a}, {\bf b}),  {\bf a} \in {\bf C}, {\bf b} \in {\bf C} \},$$  is the minimum  Hamming distance $d({\bf a}, {\bf b})$ between any two different codewords ${\bf a}$ and ${\bf b}$ in ${\bf C}$. A  $q$-ary linear code with the length $n$, the dimension $k$ (and the minimum distance $d$), is denoted by a linear $[n,k]_q$ (or $[n,k,d]_q$) code. The subcode consisting of codewords whose coordinates at one fixed position are zero, is called a shortening code. The projection code by deleting one fixed coordinate position is the punctured code. It is clear that the shortening code of a linear $[n,k,d]_q$ code is a linear $[n-1, \geq k-1, \geq d]_q$ code. The punctured code of a linear $[n, k, d]_q$ code is a linear $[n-1, k, \geq d-1]_q$ code, if $d \geq 2$. A $q$-ary code with the length $n$, the cardinality $M$, and the minimum distance $d$, is denoted by an $(n, M, d)_q$ code. The minimum distance $d$ is optimal, if $d$ is the maximal possible minimum distance of a linear $[n,k]_q$ code, and the minimum distance $d$ is best-known, if $d$ is the maximal minimum distance of all known linear $[n,k]_q$ codes. Let ${\bf C} \subset {\bf F}_q^n$ be an $(n, M, d)_q$ code, the weight distribution of ${\bf C}$ is $\Sigma_{i=0}^n A_i({\bf C})x^iy^{n-i}$, where $A_i({\bf C})$ is the number of  codewords in ${\bf C}$ with weight $i$. We refer to \cite{HP,Lint,MScode} for theory of error-correcting codes.\\

The sphere packing bound for $(n, M, d)_q$ codes asserts that $$M \cdot V_q(\lfloor \frac{d-1}{2}\rfloor) \leq q^n,$$ where $V_q(r)=1+n(q-1)+\displaystyle{n \choose 2}(q-1)^2+\cdots+\displaystyle{n \choose r}(q-1)^r$ is the volume of the ball with the radius $r$ in the Hamming metric space ${\bf F}_q^n$, since balls centered at codewords with the radius $\lfloor \frac{d-1}{2} \rfloor$ are disjoint, see \cite{HP,Lint,MScode}. Let ${\bf C} \subset {\bf F}_q^n$ be an  $(n, M, d)_q$ code, if there is no $(n, M, d+1)_q$ code, then ${\bf C}$ is called distance-optimal. Some distance-optimal codes are with respect to the sphere packing bound, namely, ${\bf C}$ is an $(n, M, d)_q$ code and $M \cdot V_q(\lfloor \frac{d}{2}\rfloor) >q^n$.\\

Cyclic codes was introduced by E. Prange in 1957  \cite{Prange}, and have been studied for many years. A linear code ${\bf C} \subset {\bf F}_q^n$
is cyclic provided that for each vector $(c_0, c_1, \ldots, c_{n-1}) \in {\bf C}$, the vector $(c_{n-1}, c_0, \ldots, c_{n-2})$ is also in ${\bf C}$.  The dual code of a cyclic code is a cyclic code. A codeword ${\bf c}$ in a cyclic code is identified with a polynomial ${\bf c}(x)=c_0+c_1x+\cdots+c_{n-1}x^{n-1}\in {\bf F}_q[x]/(x^n-1)$. Every cyclic code is a principal ideal in the ring ${\bf F}_q[x]/(x^n-1)$ and then it is generated by a factor ${\bf g}$ of $x^n-1$. The Bose-Chaudhuri-Hocquenghem (BCH) code were introduced in 1959-1960, see \cite{BC1,BC2,Hoc} for details. \\

Let $n$ be a positive integer satisfying $\gcd(n,q)=1$, and ${\bf Z}_n={\bf Z}/n{\bf Z}=\{0, 1, \ldots, n-1\}$ be the residue class module $n$. A subset $C_i$ of ${\bf Z}_n$ is called a $q$-cyclotomic coset if $$C_i=\{i, iq, \ldots, iq^{l-1}\},$$ where $i \in {\bf Z}_n$ is fixed and $l$ is the smallest positive integer such that $iq^l \equiv i$ $mod$ $n$. It is clear that $q$-cyclotomic cosets correspond to irreducible factors of $x^n-1$ in ${\bf F}_q[x]$. A generator polynomial of a cyclic code of the length $n$ is the product of several irreducible factors of $x^n-1$. Let ${\bf F}_{q^t}$ be the smallest extension field of ${\bf F}_{q}$ which contains a  primitive $n$-th root of unity $\beta$. Then the defining set of a $q$-ary cyclic code of length $n$ generated by ${\bf g}$  is the  following set $${\bf T}_{{\bf g}}=\{i: {\bf g}(\beta^i)=0\}.$$ Hence the defining set of a cyclic code is the disjoint union of several $q$-cyclotomic cosets. According to the BCH lower bound, if there are $\delta-1$ consecutive elements in the defining set, then the minimum distance of this cyclic code is at least $\delta$. $\delta$ is called the designed distance , see \cite{BC1,BC2,Hoc,HP,Lint,MScode}. There have been numerous results on BCH codes, we refer to the recent papers \cite{DingLi} and \cite{Miao}.  Some further bounds on minimum distances of cyclic codes such as the Hartmann-Tzeng bound and the Roos bound were proposed and used to construct good cyclic codes see \cite[Chapter 6]{Lint} and \cite[Chapter 4]{HP}.  Boston bounds and their generalizations for cyclic codes were given in \cite{Boston,Zeh}.\\

Negacyclic codes were introduced in 1968 by Berlekamp \cite{Berlekamp,Black}. Let $q$ be an odd prime power. A code ${\bf C} \subset {\bf F}_q^n$ is called negacyclic if $(c_0, c_1, \ldots, c_{n-1})\\ \in {\bf C}$, then $(-c_{n-1}, c_0, \ldots, c_{n-2}) \in {\bf C}$. A codeword ${\bf c}$ in a negacyclic code is identified with a polynomial ${\bf c}(x)=c_0+c_1x+\cdots+c_{n-1}x^{n-1}\in {\bf F}_q[x]/(x^n+1)$. Every negacyclic code is a principal ideal in the ring ${\bf F}_q[x]/(x^n+1)$ and then generated by a factor ${\bf g}$ of $x^n+1$.  Set ${\bf Z}_{2n}={\bf Z}/2n{\bf Z}=\{0, 1, \ldots, 2n-1\}$. A cyclotomic coset $C_i$ in ${\bf Z}_{2n}$ is called odd, if there are only odd integers in this cyclotomic coset. It is clear that odd cyclotomic cosets in ${\bf Z}_{2n}$ correspond to irreducible factors of $x^n+1$ in ${\bf F}_q[x]$. By analogy with cyclic codes, a generator polynomial of a negacyclic code is the product of several irreducible factors of $x^n+1$. The defining set of a negacyclic code generated by ${\bf g}$  is the the following set $${\bf T}_{{\bf g}}=\{i: {\bf g}(\beta^i)=0\}.$$ Then the defining set of a negacyclic code is the disjoint union of several odd cyclotomic cosets in ${\bf Z}_{2n}$. If there are $\delta-1$ consecutive odd integers in the defining set of a negacyclic code, the minimum distance of this negacyclic code is at least $\delta$. This is the BCH bound for negacyclic codes and $\delta$ is the designed distance, see \cite{Berlekamp,Black}. As pointed in \cite{SunDing}, negacyclic codes are much less studied. Several families of distance-optimal ternary negacyclic codes with small minimum distances were constructed in \cite{SunDing}. \\

\subsection{Related works and our contributions}

{\em Related works.} Many best known codes over small fields with small lengths are BCH codes, see \cite{Grassl}. For example, the optimal binary $[63, 56, 4]_2$ code and the optimal binary $[63, 50, 6]_2$ code are BCH codes. Some optimal ternary cyclic codes with the length $3^m-1$ and minimum distance four and five were constructed in \cite{LiTang}. Distance-optimal codes with minimum distance four and six were constructed in \cite{YCD,Ding,Heng,Heng1}. There are only few distance-optimal codes with minimum distance six reported in the literature, see \cite{Heng} and \cite[Theorem 9]{Wang}.\\

On the other hand, several best known codes over small fields documented in \cite{Grassl} were constructed from stored generator matrices. For example, the best known $[91, 76, 6]_3$ code and the best known $[127, 119, 4]_4$ code  were constructed from the stored generator matrices in \cite{Grassl}. The ternary $[104, 88, 6]_3$ code in \cite{Grassl} was constructed from Zinoviev code, and the $[104, 82, 8]_3$ code in \cite{Grassl} was constructed from a BCH $[121, 101, 8]_3$ code by the shortening. It is interesting and desirable if there could be cyclic, negacyclic, or Goppa codes with the same parameters as these optimal or best known codes. This would lead to fast encoding and decoding of these codes. Some best known quasi-cyclic codes were constructed in \cite{Siap,Chen1}.\\

It is a long-standing open problem that if there exists an infinite family of asymptotically good cyclic codes, see \cite{Martinez}.  Therefore it is interesting to construct infinite families of rate $\frac{1}{2}$ explicit cyclic and negacyclic codes such that their minimum distances are as large as possible. In this direction, an infinite family of binary  $[2^m-1, 2^{m-1}]_2$ cyclic codes with square-root-like lower bounds on their minimum distances and dual minimum distances was constructed in \cite{TangDing}. On the other hand, without the concerning on dual minimum distances, an infinite family of binary $[n, \frac{n+1}{2}, d\geq \frac{cn}{\log_2 n}]_2$ cyclic codes with $n=2^p-1$ and $p$ being an odd prime number, where $c$ is a positive constant, was constructed in \cite{SunLiDing}. \\

{\em Our contributions.} The main contributions of this paper are as follows.\\

1) We construct infinitely many families of distance-optimal binary BCH codes with the minimum distance $6$ and an infinite family of distance-optimal quaternary cyclic codes with minimum distance $4$.  To the best of our knowledge, these two families of distance-optimal cyclic codes are new, see \cite{YCD,Ding,Heng,Wang,Heng1}.\\

2) Several families of cyclic and negacyclic BCH codes over ${\bf F}_2$, ${\bf F}_3$, ${\bf F}_5$, ${\bf F}_7$ and ${\bf F}_9$ with good parameters $n, k, d$ such that the maximal possible minimum distance $d_{\max}$ of a  $[n, k]_q$ linear code satisfies $d_{\max} \leq d+8$, are constructed. Many codes in these families have the same parameters as best known or optimal codes in \cite{Grassl}. \\

For example, binary linear $[256-t, 225-t, 8]_2$ codes, $t=0, 1, \ldots, 84$, are constructed. They are shortening codes of a cyclic $[315, 284, 8]_2$ code. By Magma, these best known binary codes are not equivalent best known ones in \cite{Grassl}. The ternary best known  cyclic $[104, 88, 6]_3$ code, the best known  cyclic $[104, 82, 8]_3$ code and the optimal negacyclic  $[20, 14, 4]_3$ code can be obtained from our  constructions. Moreover, the optimal cyclic $[127, 119, 4]_4$ code and the best known  cyclic $[127, 112, 6]_4$ code are constructed. The  negacyclic $[91, 76, 6]_3$ code and   the negacyclic $[20,13,6]_9$ code are also constructed. Based on the Boston bound, the optimal ternary cyclic $[40, 33, 4]_3$ code is given. From Magma, all these optimal and best known codes are not equivalent to the presently best known codes in \cite{Grassl}. Many cyclic and nagecyclic $[n, k, d_{best}-1]_q$ codes are also given, where $d_{best}$ is the minimum distance of the best known $[n,k, d_{best}]_q$ code in \cite{Grassl}.\\

3) We also construct several infinite families of rate $\frac{1}{2}$ negacyclic $[n, \frac{n+1}{2}, d]_q$  or $[n, \frac{n}{2}, d]_q$ codes, and their minimum distances satisfy $d\geq \frac{cn}{\log_q n}$, where $c$ is a positive constant. These are first several families of such negacyclic codes. In such a family, a ternary negacyclic $[121,61, 22]_3$ code is given. The best known code in \cite{Grassl} is a linear $[121,61,23]_3$ code.\\

The paper is organized as follows. In Section 2, infinitely many families of distance-optimal binary BCH codes with the minimum distance six is constructed. In Section 3, an infinite family of binary BCH codes is constructed and best known  $[256-t, 225-t, 8]_2$ codes, $0 \leq t \leq 84$, are constructed as shortening codes of the binary BCH $[315, 284,8]_2$ code. In Sections 4 and 5, several families of ternary cyclic and negacyclic codes are proposed and many best known new ternary codes can be obtained from them. In Section 6, an infinite family of distance-optimal quaternary BCH codes is constructed and some new infinite families of quaternary BCH codes are constructed. In Sections 7 and 8 we present several families of negacyclic BCH codes over ${\bf F}_7$ and ${\bf F}_9$  respectively and some best known codes can be produced. In Section 9, two families of nagacyclic $[n, \frac{n+1}{2}, d\geq \frac{cn}{\log_q n}]_q$ and $[n, \frac{n}{2}, d\geq \frac{cn}{\log_q n}]_q$  codes, where $c$ is a positive constant, are presented.\\

\section{Infinitely many families of distance-optimal binary cyclic codes with the minimum distance $6$}

Let $n=3(2^m-1)$, where $m\geq3$ is an odd positive integer. It is clear that each $2$-cyclotomic coset in ${\bf Z}_n$ has at most $2m$ elements since $3|2^m+1$ and $3(2^m-1)|2^{2m}-1$. Due to  $3(2^m-1) \equiv 0$ $mod$ $n$, the $2$-cyclotomic coset $C_3$ has $m$ elements. Therefore, the defining set \begin{eqnarray}\label{binary1}{\bf T}=C_0 \bigcup C_1 \bigcup C_3\end{eqnarray} has at most $1+3m$ elements. Let $\bf{C}$ be a cyclic code generated by $\bf{T}$. Then $\bf{C}$ is a  $[3(2^m-1), 3(2^m-1)-3m-1]_2$ cyclic codes. We have the following result.\\

{\bf Theorem 2.1} {\em Let  $m\geq3$ be an odd positive integer. Then an infinite family of distance-optimal cyclic $[3(2^m-1), 3(2^m-1)-3m-1, 6]_2$ codes is constructed.}\\

{\bf Proof.} Since $0,1,2,3,4$ are in the defining set ${\bf T}$, then $d \geq 6$ from the BCH bound. Observe that the volume of the ball of the radius $3$ in the Hamming metric space ${\bf F}_2^n$ satisfies that $V_2(3)=1+n+\displaystyle{n \choose 2}+\displaystyle{n \choose 3} >\frac{n(n^2-1)}{6}>2^{3m+1}$, when $m \geq 3$. Then the codes in this family are distance-optimal.\\

The first three codes of this family have parameters $[21, 11, 6]_2$, $[93, 77, 6]_2$ and $[381, 359, 6]_2$, respectively.  From \cite{Grassl}, all of these codes are optimal codes.\\

Let the length of the binary cyclic code be $n=\frac{2^m-1}{\lambda}$, where $m=1, 2, 3, \ldots$ are positive integers and $\lambda$ is a divisor of  $2^m-1$. Then each $2$-cyclotomic coset in ${\bf Z}_n$ has at most $m$ elements. The defining set $${\bf T}=C_0 \bigcup C_1 \bigcup C_3$$ has at most $2m+1$ elements. An infinite family of cyclic $[\frac{2^m-1}{\lambda}, \geq \frac{2^m-1}{\lambda}]_2$ codes is constructed,  Therefore we have the following result.\\

{\bf Theorem 2.2} {\em If $\lambda$ is a divisor of $2^m-1$ and satisfies $\lambda<(\frac{2^{m-2}}{3(1+\epsilon)})^{1/3}$, then a binary distance-optimal cyclic $[\frac{2^m-1}{\lambda}, \frac{2^m-1}{\lambda}-2m-1,6]_2$ code is constructed, when $m$ is sufficiently large.}\\

{\bf Proof.} There are $0,1,2,3,4$ are in the defining set ${\bf T}$. Then $d \geq 6$ from the BCH bound. From the condition we get $$V_H(2,3) \geq \frac{(2^m-1)(2^m-1-\lambda)2^m-1-2\lambda)}{6\lambda^3}>(1+\epsilon)2^{2m+1},$$ when $m$ is a sufficiently large positive integer, then the minimum distance cannot be $7$ and this family of codes are distance-optimal, when $m$ is sufficiently large.\\

Notice that if $\lambda$ is fixed positive integer, $m_1, m_2, \ldots,$ are positive integers, going to the infinity, such that $\lambda|2^{m_i}-1$, then we always have the following infinite family of distance-optimal binary cyclic codes.\\

{\bf Corollary 2.1} {\em If $\lambda$ is a fixed positive integer, $m_1<m_2<m_3<, \ldots, $ is a sequence of positive integers such that $\lambda|2^{m_i}-1$, then a binary distance-optimal cyclic $[\frac{2^{m_i}-1}{\lambda}, \frac{2^{m_i}-1}{\lambda}-2m_i-1,6]_2$ code is constructed, when $m_i$ is sufficiently large.}\\

For example, an infinite family of distance-optimal cyclic $[\frac{2^{3m}-1}{7}, \frac{2^{3m}-1}{7}-6m-1, 6]_2$ codes can be obtained. Actually, infinitely many families of such distance-optimal binary cyclic codes with new parameters are constructed.\\

\section{An infinite family of binary cyclic codes}

Let $n=5(2^m-1)$, where $m=2k$ and $k\geq 1$ is an odd positive integer. Since $5|2^m+1$ and $5(2^m-1)|2^{2m}-1$,  each $2$-cyclotomic coset in ${\bf Z}_n$ has at most $2m$ elements. Besides, the $2$-cyclotomic coset $C_5$ has $m$ elements by the fact that $5(2^m-1) \equiv 0$ $mod$ $n$. Therefore, the defining set \begin{eqnarray}\label{binary2}{\bf T}=C_0 \bigcup C_1 \bigcup C_3 \bigcup C_5\end{eqnarray} has at most $5m+1$ elements. Then an infinite family of $[5(2^m-1), 5(2^m-1)-5m-1]_2$ codes is constructed, $m=3,5, \ldots$. We have the following result.\\

{\bf Theorem 3.1} {\em Let $m=2k$ and  $k\geq 1$ be an odd positive integer. Then an family of cyclic $[5(2^m-1), 5(2^m-1)-5m-1, \geq 8]_2$ codes is constructed. The maximal possible minimum distance of a linear $[5(2^m-1), 5(2^m-1)-5m-1]_2$ code satisfies $d_{\max} \leq 11$.}\\

{\bf Proof.} The first conclusion follows from the BCH bound, since the consecutive integers $0,1,2,3,4,5,6$ are in the defining set ${\bf T}$. The second conclusion follows from the sphere packing bound.\\

Let $m=6$,  a cyclic $[315, 284,8]_2$ code is obtained. Further, we can get $[256-t, 225-t, 8]_2$ shortening codes for $t=0, \ldots, 84$, from this cyclic $[315, 284,8]_2$ code. Even though these $85$ binary linear codes have the same parameters as best known ones in \cite{Grassl},  the best known linear $[256,225,8]_2$ code in \cite{Grassl} was constructed from a Goppa code. Our construction is simpler. Moreover, by Magma, we calculate weight distributions of these $85$ best known binary codes and they have different weight distributions compared with best known ones in \cite{Grassl}. For example, the best known $[256, 225, 8]_2$ code in \cite{Grassl} has $20701897$ codewords with the weight $8$. However, the shortening $[256, 225, 8]_2$ code of the binary cyclic code $[315, 284, 6]_2$ code from our construction has $405430$ codewords with the weight $8$. The shortening $[172, 141, 8]_2$ code of the binary cyclic code $[315, 284, 6]_2$ code has $14412$ codewords with the weight $8$, while that of  in \cite{Grassl} has $738177$ codewords with the weight $8$. Hence all $85$ best known codes constructed in this section are new.\\

\section{Infinite families of ternary cyclic codes}

Let $n=5(3^m-1)$,  where $m=2k$ and $k\geq 1$ is an odd positive integer. Then each $3$-cyclotomic coset in ${\bf Z}_n$ has at most $2m$ elements due to $5(3^m-1)|3^{2m}-1$, and the $3$-cyclotomic coset $C_5$ has $m$ elements due to  $5(3^m-1) \equiv 0$ $mod$ $n$. The defining set is \begin{eqnarray}\label{tenary2}{\bf T}=C_0 \bigcup C_1 \bigcup C_2 \bigcup C_4 \bigcup C_5\end{eqnarray} It has at most $7m+1$ elements. Then we have the following result.\\

{\bf Theorem 4.1} {\em Let $m=2k$ and $k\geq 1$ be an odd positive integer. Then an family of ternary cyclic $[5(3^m-1), 5(3^m-1)-7m-1, \geq 8]_3$ codes is constructed. The maximal possible minimum distance of a linear $[5(3^m-1), 5(3^m-1)-7m-1]_3$ code satisfies $d_{\max} \leq 15$.}\\

{\bf Proof.} The first conclusion follows from the BCH bound. The second conclusion follows from the sphere packing bound.\\

When $m=2$, the  linear code proposed above is reduced to be a $[40, 25, 8]_3$ code. The best known $[40, 25, 8]_3$ in \cite{Grassl} is the shortening code of a linear $[44, 29, 8]_3$ code, which was constructed from the stored generator matrix. The punctured $[39, 25, 7]_3$ code of this cyclic $[40, 25, 8]_3$ code has also the best known parameters \cite{Grassl}.\\

Observe that if the defining set  ${\bf T}=C_0 \bigcup C_1\bigcup C_5$, then the positive integers $0, 1, 3, 5$ belong to  ${\bf T}$. From the Boston bound \cite[Theorem 3]{Zeh},  an infinite family of cyclic $[5(3^m-1), 5(3^m-1)-3m-1, \geq 4]_3$ codes can be derived, where $m=2, 6, 8, \ldots$. The first code is a cyclic  $[40, 33, 4]_3$ code which is an optimal code \cite{Grassl}. Besides, shortening $[40-t, 33-t, 4]_3$ codes, $1 \leq t \leq 12$, have the same parameters as optimal codes in \cite{Grassl}.\\

Let $n=4(3^m-1)$, where $m\geq3$ is an odd positive integer. Then each $3$-cyclotomic coset in ${\bf Z}_n$ has at most $2m$ elements due to $4(3^m-1)|3^{2m}-1$, and the  $3$-cyclotomic coset $C_4$ has $m$ elements since $4(3^m-1) \equiv 0$ $mod$ $n$. Let the defining set be \begin{eqnarray}\label{tenary1}{\bf T}=C_0 \bigcup C_1 \bigcup C_2 \bigcup C_4.\end{eqnarray} Then we have the following result.\\

{\bf Theorem 4.2} {\em Let $m\geq3$ be an odd positive integer.  Then an family of ternary cyclic $[4(3^m-1), 4(3^m-1)-5m-1, \geq 6]_3$ codes is constructed. The maximal possible minimum distance of a linear $[4(3^m-1), 4(3^m-1)-5m-1]_3$ code satisfies $d_{\max} \leq 11$.}\\

{\bf Proof.} The first conclusion follows from the BCH bound. The second conclusion follows from the sphere packing bound.\\

When $m=3$, we can get a cyclic $[104, 88, 6]_3$ code. This code has the same parameter as the best known code in \cite{Grassl}. The shortening $[104-t, 88-t, 6]_3$ code for $1 \leq t \leq 11$  also has the same parameter as the best known code in \cite{Grassl}. \\

Similarly, we have the following result.\\

{\bf Theorem 4.3} {\em Let $n=4(3^m-1)$, $m\geq3$ be an odd positive integer and ${\bf T}=C_0 \bigcup C_1 \bigcup C_2 \bigcup C_4 \bigcup C_5$. Then an family of cyclic $[4(3^m-1), 4(3^m-1)-7m-1, \geq 8]_3$ codes is constructed. The maximal possible minimum distance of the $[4(3^m-1), 4(3^m-1)-7m-1]_3$ code satisfies $d_{\max} \leq 15$.}\\

{\bf Proof.} The first conclusion follows from the BCH bound. The second conclusion follows from the sphere packing bound.\\

When $m=3$, it becomes a cyclic $[104, 82, 8]_3$ code. This code has the same parameter as the best known code in \cite{Grassl} and  the shortening $[104-t, 82-t, 8]_3$ code for $1 \leq t \leq 12$ also has the same parameter as the best known code in \cite{Grassl}. \\

\section{Infinite families of negacyclic ternary codes}
Let $n=5 \cdot \frac{3^m-1}{2}$, where $m=2k$ and $k$ is an odd positive integer. It is clear that each $3$-cyclotomic coset in ${\bf Z}_{2n}$ has at most $2m$ elements and the $3$-cyclotomic coset $C_{5}$ has $m$ elements. Let
 defining set be $${\bf T}=C_1 \bigcup C_5.$$ Then there is at most $3m$ elements in ${\bf T}$. Therefore, an infinite family of negacyclic codes can be obtained.\\

{\bf Theorem 5.1} {\em Let $m=2k$ and $k$ be an odd positive integer.  The  codes defined as above are negacyclic $[5\cdot \frac{3^m-1}{2}, 5\cdot \frac{3^m-1}{2}-3m, \geq 4]_3$ codes. The maximal possible minimum distance satisfies $d_{\max}(5\cdot \frac{3^m-1}{2}, 5\cdot \frac{3^m-1}{2}-3m) \leq 7$. }\\

{\bf Proof.} The first conclusion is proved in the construction. The second conclusion follows from the sphere packing bound immediately.\\

Let $m=2$, then the code in Theorem 5.1 is a negacyclic $[20, 14, 4]_3$ code which is optimal \cite{Grassl}. By Magma, there are only $120$  codewords with weight $4$ in this optimal $[20,14,4]_3$ negacyclic code.  However, one can check that the $[20,14,4]_3$ code documented in \cite{Grassl} is a subcode of the Plotkin sum $[20,15,4]_3$ code, which has much more codewords with weight $4$. Therefore these two $[20,14,4]_3$ codes are not equivalent with respect to the monomial action. From Magma,  the weight distribution of this negacyclic $[20, 14, 4]_3$ code is $y^{20}+120x^4y^{16}+720x^5y^{15}+3360x^6y^{14}+13440x^7y^{13}+43920x^8y^{12}+121000x^9y^{11}+256080x^{10}y^{10}
+4651250x^{11}y^9+726120x^{12}y^8+860640x^{13}y^7+86110x^{14}y^6+712608x^{15}y^5+430440x^{16}y^4+202320x^{17}y^3
+70600x^{18}y^2+13920x^{19}y+71440x^{20}$. The dual code of this optimal negacyclic $[20,14,4]_3$ code is a negacyclic $[20, 6, 9]_3$ code. The corresponding best known code in \cite{Grassl} is a ternary $[20,6,10]_3$ code from the stored generator matrix.\\

By the shortening, the optimal $[15, 9, 4]_3$ code, the optimal $[16, 10, 4]_3$ code, the optimal $[17, 11, 4]_3$ code, the optimal $[18, 12, 4]_3$ code and the optimal $[19, 13, 4]_3$ can also be obtained from this optimal negacyclic $[20, 14, 4]_3$ code. From the Y1 construction of \cite[Lemma 3.4]{Grassl1}, we can also get a  $[16, 3, 9]_3$ linear code. The corresponding best known code in \cite{Grassl} is a linear $[16,3,10]_3$ code.\\

By using the defining set $${\bf T_1}=C_1 \bigcup C_5 \bigcup C_7$$  and $${\bf T_2}=C_{-5} \bigcup C_{-1} \bigcup C_1 \bigcup C_5,$$ we have the result as below.\\

{\bf Theorem 5.2} {\em Let $m=2k$, where $k$ is an odd positive integer and $n=5 \cdot \frac{3^m-1}{2}$.  An infinite family of negacyclic $[5\cdot \frac{3^m-1}{2}, 5\cdot \frac{3^m-1}{2}-5m, \geq 6]_3$ codes and an infinite family of negacyclic $[5\cdot \frac{3^m-1}{2}, 5\cdot \frac{3^m-1}{2}-6m, \geq 7]_3$ codes are constructed. }\\

Note that when $m=2$, the negacyclic $[20, 10, 6]_3$ code and the negacuclic $[20, 8, 7]_3$ code are constructed. It is known that the  best known ones in \cite{Grassl} is the $[20, 10, 7]_3$ linear code and the $[20, 8, 9]_3$ linear code, respectively.\\

%

Let $n=7 \cdot \frac{3^m-1}{2}$, where $m=3k$ and  $k$ is an odd positive integer. Since $3^3+1|3^{3k}+1$ and $7 (3^m-1) \equiv 0$ $mod$ $2n$, there are at most $2m$ elements in each $3$-cyclotomic coset of ${\bf Z}_{2n}$ and there are $m$ elements in the $3$-cyclotomic coset $C_7$. Let the defining set be $${\bf T}=C_1 \bigcup C_5 \bigcup C_7.$$ Since $1,3,5,7,9$ belong to ${\bf T}$, we have the following conclusion. \\

{\bf Theorem 5.3} {\em Let $m$ be a positive integer of the form $m=3k$, where $k$ is an odd positive integer. The infinite family of negacyclic $[7 \cdot \frac{3^m-1}{2}, 7 \cdot \frac{3^m-1}{2}-5m, \geq 6]_3$ codes with the defining set ${\bf T}$ is constructed.  The maximal possible minimum distance satisfies $d_{\max}(7 \cdot \frac{3^m-1}{2}, 7 \cdot \frac{3^m-1}{2}-5m) \leq 11$.}\\

Note that the first code in this family is a negacyclic  $[91, 76, 6]_3$ code, which  has the best known minimum distance as in \cite{Grassl}. The best known $[91,76,6]_3$ code documented in \cite{Grassl} is from a stored generator matrix, while the  ternary negacyclic code from our construction has the algebraic structure. On the other hand, from Magma,  this negacyclic $[91, 76, 6]_3$ code in Theorem 6.3 is not equivalent to that  in \cite{Grassl}.
Moreover, the best known $[90, 75, 6]_3$ code, $[89, 74, 6]_3$ code, $[88, 73, 6]_3$ code, $[87, 72, 6]_3$ code and $[86, 71, 6]_3$ code can be constructed by shortening  the negacyclic $[91, 76, 6]_3$ code. \\

We recall the following lemma in \cite{LDL}, which is useful to determine sizes of cyclotomic cosets.\\

{\bf Lemma 5.1 (see \cite{LDL})} {\em Let $q$ be a prime power, $a$ and $b$ be two positive integers. Then $\gcd(q^a-1, q^b-1)=q^{\gcd(a,b)}-1$.}\\

Consider  ternary negacyclic codes of the length $n=\frac{3^p-1}{2}$, where $p$ is prime. Since $n=3^{p-1}+3^{p-2}+\cdots+3+1$ is odd,  the following result follows from Lemma 6.1 immediately.\\

{\bf Lemma 5.2} {\em An odd $3$-cyclotomic coset $C_i$ in ${\bf Z}_{2n}$ has $p$ elements if $i \neq 0$ and $i \neq n$.}\\

{\bf Proof.} From the fact $i$ is odd and $\gcd(3^b-1, 3^p-1)=3^{\gcd(b,p)}-1$, one can get that  that if $i(3^b-1) \equiv 0$ $mod$ $3^p-1$,  then  $b=p$.\\

Let us observe the following example to see how to get a lower bound on the minimum distance of a negacyclic $[\frac{3^p-1}{2}, \frac{3^p+1}{4}]_3$ code.\\

{\bf Example 5.1.} Set $p=5$, then $n=\frac{3^p-1}{2}=121$. The defining set ${\bf T}$ is the disjoint union of the following first $12$ $3$-cyclotomic cosets with $5$ elements in ${\bf Z}/242{\bf Z}$.\\

$C_1=\{1,3,9,27, 81\}$, $C_5=\{5,15,45,135, 163\}$, $C_7=\{7,21,63,83,189\}$, $C_{11}=\{11,33,\\55,99,165\}$, $C_{13}=\{13, 39,85,109,117\}$, $C_{17}=\{17,51,153,167,217\}$, $C_{19}=\{19,29,57,87,\\171\}$  $C_{23}=\{23,69,137,169,207\}$, $C_{25}=\{25,75,89,191,225\}$, $C_{31}=\{31,37,91,93,111\}$, $C_{35}=\{35,73,105,173,219\}$, $C_{41}=\{41,123,127,139,175\}$. There are $21$ consecutive odd positive integers $1,3,5,\ldots, 41$ in this defining set, and $45$ is not in this defining set. Therefore, we can construct a negacyclic $[121, 61, 22]_3$ code and the  best known one in \cite{Grassl} is a linear $[121, 61, 23]_3$ code.\\

The positive integer $\frac{n-1}{2}=\frac{3^p-3}{4}$ satisfies $\frac{n-1}{2} =\frac{3^p-3}{4} \equiv 0$ $mod$ $p$ from the Fermat litter theorem. As in Example 5.1, we take the union of first $\frac{3^p-3}{4p}$ odd $3$-cyclotomic cosets with $p$ elements as the defining set ${\bf T}$. Then a family of negacyclic $[\frac{3^p-1}{2}, \frac{3^p+1}{4}]_3$ code. We have the following result.\\

{\bf Theorem 5.4} {\em Let $p=3,5,7,11, \ldots$ be prime positive integers. An infinite family of negacyclic $[\frac{3^p-1}{2}, \frac{3^p+1}{4}, \geq \frac{3^p-3}{4p}]_3$ is constructed.}\\

{\bf Proof.} There are at least $\frac{3^p-3}{4p}$ consecutive odd positive integers in the defining set. The conclusion is proved.\\

Observe that the there are actually more than $\frac{3^p-3}{4p}$ consecutive odd positive integers in the defining set, since some odd positive integers occur in the previous $3$-cyclotomic cosets, as in Example 5.1. The real BCH lower bound in Example 5.1 is $d\geq 22$, the lower bound in Theorem 5.4, when $p=5$, is $d \geq \frac{240}{20}=12$. Therefore it is reasonable to conjecture that the lower bound on the minimum distance in Theorem 5.4 can be improved significantly.\\

\section{Infinite families of quaternary cyclic codes}

%

Let $n=2^{2m-1}-1$, where $m$ is a positive integer. Note that $4^m-2=2(2^{2m-1}-1) \equiv 0$ $mod$ $n$. Thus one has $2\in C_1$. Let the defining set be ${\bf T}=C_0 \bigcup C_1$. We get an infinite family of distance-optimal quaternary cyclic codes with minimum distance $4$.\\

{\bf Theorem 6.1} {\em Let $m$ be a positive integer satisfying $m \geq 3$. The codes defined above is a family of  distance-optimal cyclic $[2^{2m-1}-1, 2^{2m-1}-1-2m, 4]_4$ codes.}\\

{\bf Proof.} From the BCH bound, the minimum distance of codes in the above family is at least $4$. It is clear that $$V_4(2) \geq \frac{9(2^{2m-1}-1)(2^{2m-1}-2)}{2} \geq 4^{2m}.$$ Then these codes are distance-optimal with respect to the sphere packing bounds.\\

Notice these distance-optimal quaternary cyclic codes can be compared with distance-optimal quaternary codes in \cite{Heng1}. The first three codes in this family have parameters $[31, 25, 4]_4$, $[127, 119, 4]_4$ and $[511, 501, 4]_4$. The shortening  of the $[511, 501, 4]_4$ code is a  $[256, 246, 4]_4$ linear code. The best known code in \cite{Grassl} is a linear $[256, 246, 5]_4$ code.\\

By choosing other defining set, we can also obtain infinite families of quaternary cyclic codes with minimum distance $6,\,8$ and $10$.\\

{\bf Theorem 6.2} {\em Let $m$ be a positive integer satisfying $m \geq 3$ and $n=2^{2m-1}-1$.

{\rm (1)} When ${\bf T}=C_0 \bigcup C_1 \bigcup C_3$, the code with the defining set ${\bf T}$ is a cyclic $[2^{2m-1}-1, 2^{2m-1}-4m, \geq 6]_4$ code;

{\rm (2)} When ${\bf T}=C_0 \bigcup C_1 \bigcup C_3 \bigcup C_5$, the code with the defining set ${\bf T}$ is a cyclic $[2^{2m-1}-1,2^{2m-1}-6m+1, \geq 8]_4$ code;

{\rm (3)} When ${\bf T}=C_0 \bigcup C_1 \bigcup C_3 \bigcup C_5 \bigcup C_7$, the code with the defining set ${\bf T}$ is a cyclic $[2^{2m-1}-1,2^{2m-1}-8m+2, \geq 10]_4$ code.}\\

If $m=3,\,4,\,5$, then the corresponding codes of Theorem 5.2 (1)  have parameters  $[31,20,6]_4$,  $[127, 112, 6]_4$ and $[511, 492, 6]_4$. It can be checked that there is a linear  $[31, 20, 7]_4$ code listed in \cite{Grassl} and the quaternary cyclic $[127, 112, 6]_4$ code has the same parameters as best known quaternary code \cite{Grassl}. By shortening the $[511, 492, 6]_4$ code, one can get a linear $[256, 237, 6]_4$ code and the best known code is a linear $[256, 237, 7]_4$ code \cite{Grassl}. If $m=4$, then codes in Theorem 5.2 (2) and (3) become the cyclic $[127, 105, 8]_4$ code and the cyclic $[127, 98, 10]_4$ code. The best known ones in \cite{Grassl} are $[127, 105, 9]_4$ and $[127, 98,11]_4$  codes, respectively.

%
%
%
%
%
%
%
%
%
%
%
%
%

\section{Infinite families of negacyclic codes over ${\bf F}_7$}

Let $n=5 \cdot \frac{7^m-1}{6}$, or $5 \cdot \frac{7^m-1}{8}$ or $5 \cdot \frac{7^m-1}{4}$, where $m=2k$ and $k$ is an odd positive integer. Since $7^2+1|7^m+1$ and $5 \cdot (7^m-1) \equiv 0$ $mod$ $2n$,  each $7$-cyclotomic coset in ${\bf Z}_{2n}$ has at most $2m$ elements and the $7$-cyclotomic coset $C_5$ has $m$ elements. Define the defining set $${\bf T}=C_1 \bigcup C_3 \bigcup C_5.$$

{\bf Theorem 7.1} {\em Let $n$ and ${\bf T}$ be defined as above. An infinite family of negacyclic $[n, n-5m, \geq 5]_7$ codes is constructed. The  maximal possible minimum distance is $d_{\max}(n, n-5m) \leq 11$.}\\

{\bf Proof.}  The conclusion follows from the BCH bound for negacyclic codes. The second conclusion follows from the sphere packing bound.\\

When $m=2$,  we can obtain the negacyclic $[30, 20, 5]_7$ code, the  negacyclic $[40, 30, 5]_7$ code and the negacyclic $[60, 50, 5]_7$ code from  Theorem 7.1. The corresponding best known codes in \cite{Grassl} are the linear $[30, 20, 7]_7$ code,  the linear $[40, 30, 7]_7$ code and the linear $[60, 50, 6]_7$  code.\\

\section{An infinite family of negacyclic codes over ${\bf F}_9$}

Let $n=5\frac{9^m-1}{2}$, where $m=1,3,5,\ldots$ are odd positive integers. It is clear that there are at most $2m$ elements in each $9$-cyclotomic coset of ${\bf Z}_{2n}$ and there are $m$ elements in the $9$-cyclotomic coset $C_5$. Let  $${\bf T}=C_1 \bigcup C_3 \bigcup C_5 \bigcup C_7.$$ Then we have the following result.\\

{\bf Theorem 8.1} {\em Let $m$ be an odd positive integer. An infinite family of negacyclic $[5\frac{9^m-1}{2}, 5\frac{9^m-1}{2}-7m, \geq 6]_9$  codes is constructed. The maximal possible distance is $d_{\max}(5\frac{9^m-1}{2}, 5\frac{9^m-1}{2}-7m)\leq 15$.}\\

{\bf Proof.} Since consecutive odd positive integers $1,3,5,7,9$ are in the defining set ${\bf T}$, the first conclusion follows from the BCH bound for negacyclic codes. The second conclusion follows from the sphere packing bound.\\

The first code in this family is the negacyclic $[20, 13, 6]_9$ code. The corresponding best known linear $[20,13,6]_9$ code documented in \cite{Grassl} was constructed from a stored generator matrix. Moreover, the best known $[19,12,6]_9$ code, $[18, 11, 6]_9$ code and $[19,13,5]_9$ code can be obtained from the $[20, 13, 6]_9$ code in this paper by the shortening and the puncture.\\

\section{Infinite families of negacyclic $[n, \frac{n+1}{2},d]_q$ or $[n, \frac{n}{2},d]_q$ codes with $d \geq \frac{cn}{\log_q n}$}

Let $q$ be an odd prime power satisfying $q \equiv 3$ $mod$ $4$. Then we consider negacyclic $q$-ary codes of the length $n=\frac{q^p-1}{q-1}$, where $p$ is prime. It is clear that the length $n=q^{p-1}+q^{p-2}+\cdots+q+1$ is odd.\\

{\bf Lemma 9.1} {\em An odd $q$-cyclotomic coset $C_i$ in ${\bf Z}_{2n}$ has $p$ elements if $i \neq 0$ and $i \neq n$.}\\

{\bf Proof.} From the fact $i$ is odd and  $\gcd(q^b-1, q^p-1)=q^{\gcd(b,p)}-1$,  one can get that  that if  $i(q^b-1) \equiv 0$ $mod$ $\frac{2(q^p-1)}{q-1}$, then $b=p$.\\

The positive integer $\frac{n-1}{2}=\frac{q^p-q}{2(q-1)}$ satisfies $\frac{n-1}{2} \equiv 0$ $mod$ $p$ from the Fermat litter Theorem. Since there are only two odd $q$-cyclotomic cosets $C_0$ and $C_n$ with one elements. We can take the union of the first $\frac{q^p-q}{2(q-1)p}$ odd $q$-cyclotomic cosets as the defining set ${\bf T}$. There are exactly $p \cdot \frac{q^p-q}{2(q-1)p}=\frac{q^p-q}{2}=\frac{n-1}{2}$ elements in ${\bf T}$. We construct an infinite family of negacyclic $[n, \frac{n+1}{2}]_q$ codes. We have the following lower bound on their minimum distances.\\

{\bf Theorem 9.1} {\em Let $q$ be an odd prime power satisfying $q \equiv 3$ $mod$ $4$ and $p=3,5,7,11, \ldots$ be prime positive integers. An infinite family of  negacyclic $[\frac{q^p-1}{q-1}, \frac{q^p+q-2}{2(q-1)}, \geq \frac{q^p-q}{2(q-1)p}]_q$ is constructed.}\\

{\bf Proof.} There are at least $\frac{q^p-q}{2(q-1)p}$ consecutive odd positive integers in this defining set. Then the conclusion is proved.\\

When $q=3$, the above infinite family of  $q$-ary negacyclic codes is the same as the family of  ternary negacyclic codes constructed in Theorem 5.4. As showed in Example 5.1, the real BCH lower bound can be larger, since some odd positive integers occur in the previous cyclotomic cosets.\\

Now we consider the negacyclic codes with the length $n=\frac{q^p-1}{2}$, where $q$ is an odd prime power satisfying $q \equiv 1$ $mod$ $4$, and $p=1,3,5,7,11,\ldots$ are prime positive integers. The following lemma about $q$-cyclotomic cosets in ${\bf Z}_{2n}$ is essential in our construction.\\

{\bf Lemma 9.2} {\em Let $q$ is an odd prime power, $p$ be an odd prime integer and $n=\frac{q^p-1}{2}$. Each odd $q$-cyclotomic coset $C_i$ in ${\bf Z}_{2n}$ has $p$ elements if $i$ is not a multiple of $\frac{q^p-1}{q-1}$. When $i \equiv 0$ $mod$ $\frac{q^p-1}{q-1}$, the odd $q$-cyclotomic coset $C_i$ has exactly one element. Moreover, there are $\frac{q-1}{2}$ odd $q$-cyclotomic cosets in ${\bf Z}_{2n}$ with one element.}\\

{\bf Proof.} It is clear that each $q$-cyclotomic coset has at most $p$ elements. If $i(q^b-1) \equiv 0$ $mod$ $q^p-1$ and $b\ne p$, then $\gcd(q^b-1, q^p-1)=q-1$ and $i \equiv 0$ $mod$ $\frac{q^p-1}{q-1}$. The other conclusions follow directly.\\

Let us observe the following example.\\

{\bf Example 9.1.} Set $q=5$, $p=3$, then $n=\frac{q^p-1}{2}=62$. The defining set ${\bf T}$  is the disjoint union of the following odd $5$-cyclotomic cosets, $C_1=\{1, 5, 25\}$,
$C_3=\{3, 15, 75\}$, $C_7=\{7, 35, 51\}$, $C_9=\{9, 45, 101\}$,
$C_{11}=\{11, 27, 55\}$, $C_{13}=\{13, 65, 77\}$, $C_{17}=\{17, 53, 85\}$, $C_{19}=\{19, 95, 103\}$, $C_{21}=\{21, 29, 105\}$, $C_{23}=\{23, 79, 115\}$, $C_{31}=\{31\}$. There are $16$ consecutive odd positive integers $1, 3, \ldots, 31$ in this defining set. Therefore, we get a negacyclic  $[62, 30, 17]_5$ code from the BCH bound. The  corresponding best known code in \cite{Grassl} is a linear $[62, 30, 20]_5$  code. The minimum distance $17$ is confirmed by Magma.\\

Notice that $\frac{n}{2}=\frac{q^p-1}{4}=\frac{q^p-q}{4}+\frac{q-1}{4}$. The term $\frac{q^p-q}{4}$ is a multiple of the odd prime positive integer $p$ from the Fermat little theorem. Then we can take the defining set ${\bf T}$ as the union of the first $\frac{q^p-q}{4p}$ odd $q$-cyclotomic cosets with $p$ elements and the first $\frac{q-1}{4}$ odd $q$-cyclotomic cosets with one element. There are $\frac{n}{2}$ elements in this defining set. Then a family of negacyclic $[n, \frac{n}{2}]_q$ codes is constructed.  We have the following lower bounds on their minimum distances.\\

{\bf Theorem 9.2} {\em Let $q$ be an odd prime power satisfying $q \equiv 1$ $mod$ $4$, where $p=1,3,5,7,11,\ldots$ are prime positive integers. An infinite family of negacyclic $[\frac{q^p-1}{2}, \frac{q^p-1}{4}, d\geq \frac{q^p-q}{4p}]_q$ codes is constructed.}\\

{\bf Proof.} There are at least $\frac{q^p-q}{4p}$ consecutive odd positive integers in this defining set. The conclusion is proved.\\

Notice that the real BCH lower bound $d \geq 17$ in Example 9.1 is better than $d \geq 10$ in Theorem 9.2 for $q=5$, $p=3$. Actually, some odd positive integers occur in previous $q$-cyclotomic cosets. It is reasonable to conjecture that lower bounds in Theorem 9.1 and 9.2 can be improved significantly.\\

\section{Conclusions}

In this paper, infinite many families of distance-optimal binary cyclic codes with the minimum distance $6$ and an infinite family of quaternary cyclic codes with the minimum distance $4$ were constructed. Several infinite families of  negacyclic $[n, \frac{n+1}{2}, d\geq \frac{cn}{\log_qn}]_q$ codes or negacyclic $[n, \frac{n}{2}, d \geq \frac{cn}{\log_qn}]_q$ codes were also presented. Then $145$ optimal or best known codes as cyclic codes, negacyclic codes, their shortening codes and their punctured codes can be produced from our constructions. From Magma, all optimal or best known codes constructed in this paper are not equivalent to the presently known codes in \cite{Grassl}. We list optimal or best known codes constructed in this paper in Table 1. In Table 2, we list some codes with minimum distances $d_{best}-1$, which were constructed in this paper .\\

\begin{longtable}{|l|l|l|}
\caption{\label{tab:A-q-5-3} Cyclic and negacyclic codes, shortening and punctured codes.}\\ \hline
$q$&Codes in this paper&Best known or optimal codes \\ \hline
$2$ &$[21, 11, 6]_2$& optimal \\ \hline
$2$ &$[93, 77, 6]_2$& optimal  \\ \hline
$4$ &$[31, 25, 4]_4$& optimal  \\ \hline
$4$ &$[127, 119, 4]_4$& optimal  \\ \hline
$2$ &$[256-t, 225-t, 8]_2, 0 \leq t \leq 84$ new&$[256-t, 225-t, 8]_2, 0 \leq t \leq 84$ \\ \hline
$3$ &$[40-t, 33-t, 4]_3, 0\leq t \leq 12$ new& $[40-t, 33-t, 4]_3, 0\leq t \leq 12$ \\ \hline
$3$ &$[40-t, 25, 8-t]_3$, $0 \leq t \leq 1$ new& $[40-t, 25, 8-t]_3$, $0 \leq t \leq 1$ \\ \hline
$3$ &$[104-t, 88-t, 6]_3, 0 \leq t \leq 11$ new& $[104-t, 88-t, 6]_3, 0 \leq t \leq 11$ \\ \hline
$3$ &$[104-t, 82-t, 8]_3, 0 \leq t \leq 12$ new&$[104-t, 82-t, 8]_3, 0 \leq t \leq 12$ \\ \hline
$3$ &$[20-t,14-t,4]_3, 0 \leq t \leq 5$ new&$[20-t,14-t,4]_3, 0 \leq t \leq 5$ \\ \hline
$3$ &$[91-t,76-t,6]_3, 0 \leq t \leq 5$ new&$[91-t,76-t,6]_3, 0 \leq t \leq 5$ \\ \hline
$4$ &$[127-t, 112-t, 6]_4$, $0 \leq t \leq 4$ new& $[127, 112-t, 6]_4$, $0 \leq t \leq 4$ \\ \hline
$9$ &$[20-t,13-t,6]_9, 0 \leq t \leq 2$ new&$[20-t,13-t,6]_9, 0 \leq t \leq 2$ \\ \hline
$9$ &$[19,13,5]_9$ new&$[19,13,5]_9$ \\ \hline
\end{longtable}

\begin{longtable}{|l|l|l|}
\caption{\label{tab:A-q-5-3} Cyclic and negacyclic codes, shortening and punctured codes with $d=d_{best}-1$.}\\ \hline
$q$&Codes in this paper&Best known codes \\ \hline
$3$ &$[16, 3, 9]_3$&$[16,3, 10]_3$ \\ \hline
$3$ &$[20, 10, 6]_3$& $[20,10,7]_3$ \\ \hline
$3$ &$[20, 6, 9]_3$& $[20, 6, 10]_3$ \\ \hline
$3$ &$[121,61,22]_3$ & $[121,61,23]_3$  \\ \hline
$4$ &$[30, 20, 6]_4$ & $[30, 20, 7]_4$ \\ \hline
$4$ &$[127, 98, 10]_4$ &$[127, 98, 11]_4$ \\ \hline
$4$ &$[127, 105, 8]_4$ &$[127, 105, 9]_4$ \\ \hline
$4$ &$[256, 246, 4]_4$ &$[256, 246, 5]_4$ \\ \hline
$4$ &$[256, 237, 6]_4$ &$[256, 237, 7]_4$ \\ \hline
$7$ &$[60, 50, 5]_7$&$[60, 50, 6]_7$ \\ \hline
\end{longtable}


\begin{thebibliography}{10}

\bibitem{Berlekamp} E. R. Berlekamp, Negacyclic codes for the Lee metric, Proc. Conf. Combin. Math. and Appl., Chapel Hill, NC, pp. 298-316, 1968.


\bibitem{Black} T. Blackford, Negacyclic duadic codes, Finite Fields Appl., vol. 14, no. 4, pp. 930-943, 2008.


\bibitem{BC1} R. C. Bose and D. K. Ray-Chaudhuri, On a class of error-correcting binary group codes, Inf. and Contr., vol. 3, pp. 68-79, 1960.

\bibitem{BC2} R. C. Bose and D. K. Ray-Chaudhuri, Further results on error-correcting binary group codes, Inf. and Contr., vol. 3, pp. 279-290, 1960.

\bibitem{Boston} N. Boston, Bounding minimum distances of cyclic codes using algebraic geometry, Electron. Notes Discr. Math., vol. 6, pp. 385-394, 2001.



\bibitem{Chen1} E. Z. Chen, New quasi-cyclic codes from simplex codes, IEEE Trans. Inf. Theory, vol. 53,  no. 3, pp. 1193-1196, 2007.


\bibitem{Ding} C. Ding and T. Helleseth, Optimal ternary cyclic codes from monomials, IEEE Trans. Inf. Theory, vol. 59, no. 9, pp. 5898-5904, 2013.

\bibitem{DingLi} C. Ding and C. Li, BCH cyclic codes, submitted, 2023.



\bibitem{Grassl} M. Grassl, Bounds on the minimum distance of linear codes and quantum codes,
Online available at http://www.codetables.de.

\bibitem{Grassl1} M. Grassl, Searching for linear codes with large minimum distances, In: W. Bosma and J. Cannon (Eds.), Discovering Mathematics with Magma, Springer, Berlin, Heidelberg, Tokyo, 2006.


\bibitem{Heng} Z. Heng, C. Ding and W. Wang, Optimal binary linear codes from maximal arcs, IEEE Trans. Inf. Theory, vol. 66, no. 9, pp. 5387-5394, 2020.




\bibitem{Heng1} Z. Heng, Q. Wang and C. Ding, Two families of optimal linear code and their subfield codes, IEEE Trans. Inf. Theory, vol. 66, no. 11, pp. 6872-6883, 2020.


\bibitem{Hoc} A. Hocquenghem, Codes correcteurs d'erreurs, Chiffres (Paris), vol. 2, pp. 147-156, 1959.



\bibitem{HP} W. C. Huffman and V. Pless, Fundamentals of error-correcting codes, Cambridge University Press, Cambridge, U. K., 2003.


\bibitem{LDL} C. Li, C. Ding and S. Li, LCD cyclic codes over finite fields, IEEE Trans. Inf. Theory, vol. 63, no. 7, pp. 4344-4356, 2017.



\bibitem{LiTang} N. Li, C. Li, T. Helleseth, C. Ding and X. Tang, Optimal ternary cyclic codes with minimum distance four and five, Finite Fields Appl., vol. 30, pp. 100-120, 2014.




\bibitem{MScode} F. J.  MacWilliams and N. J. A. Sloane, The theory of error-correcting codes, 3rd Edition, North-Holland Mathematical Library, vol. 16. North-Holland, Amsterdam, 1977.



\bibitem{Martinez} C. Mart\'{i}nez-P\'{e}rez and W. Willems, Is the class of cyclic codes asymptotically good? IEEE Trans. Inf. Theory, vol. 52, no. 2, pp. 696-700, 2006.

\bibitem{Miao} S. Noguchi, X-N. Lu, M. Jimbo and Y. Miao, BCH codes with minimum distances propotional to code lengths, SIAM J. Discr. Math., vol. 35, no. 1, pp. 179-193, 2021.

\bibitem{Prange} E. Prange, Cyclic error-correcting codes in two symbols, TN-57-013, Technical notes issued by Air Force Cambridge Research Labs, 1957.

\bibitem{Siap} I. Siap, N. Aydin and D. K. Ray-Chaydhuri, New ternary quasi-cyclic codes with better minimum distances, IEEE Trans. Inf. Theory, vol. 46, no. 4, pp. 1554-1558, 2000.

\bibitem{SunDing} Z. Sun and C. Ding, Several familes of ternary negacyclic codes and their duals, early access in IEEE Trans. Inf. Theory.

\bibitem{SunLiDing} Z. Sun, C. Li and C. Ding, An infinite family of binary cyclic codes with best parameters, early access in IEEE Trans. Inf. Theory.

\bibitem{TangDing} C. Tang and C. Ding, Binary $[n, \frac{n+1}{2}]$ cyclic codes with good minimum distances, IEEE Trans. Inf. Theory, vol. 68, no. 12, pp. 7842-7849, 2022.



\bibitem{Lint} J. H. van Lint, Introduction to the coding theory,  Third and Expanded Edition, vol. 86,  Springer, Berlin, 1999.

\bibitem{Wang} X. Wang, D. Zheng and C. Ding, Some punctured codes of sveral families of binary linear codes, IEEE Trans. Inf. Theory, vol. 67, no. 8, pp. 5133-5148, 2021.

\bibitem{YCD} J. Yuan, C. Carlet and C. Ding, The weight distribution of a class
of linear codes from perfect nonlinear functions, IEEE Trans. Inf.
Theory, vol. 52, no. 2, pp. 712-717, 2006.

\bibitem{Zeh} A. Zeh, A. Wachter-Zeh and S. Bezzateev, Decoding cyclic codes up to a new bound on the minimum distance, IEEE Trans. Inf. Theory, vol. 58, no. 6, pp. 3951-3960, 2012.





\end{thebibliography}
\end{document}